\documentclass[12pt,preprint2]{aastex}
\newcommand{\etal}{et al.~}
\newcommand{\starA} {Star~A }

\begin{document}

\title{The bright optical companion to the eclipsing millisecond
pulsar in NGC 6397\footnote{Based on
observations with the NASA/ESA HST, obtained at
the Space Telescope Science Institute, which is operated by AURA, Inc.,
under NASA contract NAS5-26555}} 

\author{Francesco R. Ferraro, Andrea Possenti, Nichi D'Amico,
Elena Sabbi}
\affil{Osservatorio Astronomico di Bologna, via Ranzani 1,I--40126 Bologna, 
Italy \\
$~$\vskip 1.0truecm
Received by ApJ Letters on 31 July, 2001}

\begin{abstract}
We report  the possible optical identification of the companion to 
the eclipsing millisecond pulsar PSR~J1740$-$5340 in the globular 
cluster NGC 6397. A bright variable star with an anomalous red 
colour and optical variability which nicely correlates to the orbital 
period of the pulsar has been found close to the pulsar position.  
If confirmed, the optical light curve, reminiscent of tidal 
distorsions similar to those observed in detached and contact 
binaries, support the idea that this is the first case of a Roche 
lobe filling companion to a millisecond pulsar.
\end{abstract}


\section{Introduction} 
\label{sec:intro}

The millisecond pulsar (MSP) PSR J1740$-$5340 was discovered 
during a systematic search of the globular cluster (GC) system 
for millisecond pulsars, 
carried out with the Parkes radiotelescope (D'Amico \etal 2001a, 2001b).  
The pulsar, associated with the globular cluster NGC6397, is 
member of a binary system with a 
relatively wide orbit of period $\simeq$ 1.35 days, and it is eclipsed
for about 40\% of the orbital phase at 1.4 GHz. In a companion paper,
D'Amico et al. (2001c)\nocite{dpms+01} provide strong evidences that the
companion star could be an unusual object, and give a precise
position for the pulsar. 
 
We here present the results of a deep search of the optical companion 
to PSR J1740$-$5340 in the HST archive.
 
\section{Observations and  data analysis}
\label{sec:obs}

The photometric data consists of a  series of public
HST exposures  taken on March 1996
and April 1999, retrieved
from the   {\it  ESO/ST-ECF Science Archive}. The 1999 observations 
consist of
116 WFPC2 exposures with filters F555W, F675W, F814W and F656N
(referred here as $V_{555}, R_{675}, I_{814}$ and H$\alpha$),
spanning about 1.8 days.  The 1996 observations  consist of 
55 exposures with filters F336W and F439W (referred here as 
$U_{336}$ and $B_{439}$), spanning 0.4 and 0.2 days respectively.

From the accurate timing position (RAJ 17$^{\rm h}$ 40$^{\rm m}$ 44\fs589;
DECJ $-$53$^{o}$ 40$\arcmin$ 40$\farcs9$) the  
MSP in NGC6397 turns out to be 
approximately at $29''E $ and $16''S$ from the cluster
center (Djorgovski \& Meylan 1993). We used the STSDAS program $metric$, 
to roughly locate the MSP in the retrieved WFPC2 images, and it turns out 
to be within the field of view of the WF4 chip in both the data set.  

\subsection{Astrometry}

The exact location of the MSP in the HST images was obtained
by searching an astrometric solution for a wide field CCD image
of a region around NGC 6397. We retrieved from the {\it ESO Science Archive}
an image obtained on May 1999 with the Wide Field Imager (WFI)
at the ESO 2.2m telescope (at European Souther Observatory,  La Silla, Chile).
The entire image consists of a mosaic of 8 chips (each with a field of
view of  $8'\times 16'$) giving a global  field of view of
$33'\times 34'$. Only the chip containing the cluster center was used. 
The new astrometric {\it Guide Star Catalog} ($GSCII$) recently released and  
now available on the WEB, was used to search for astrometric standard
lying in the WFI image field of view: several hundreds astrometric 
$GSCII$ reference stars have been found, 
allowing an accurate absolute positioning of the image.
     
In order to derive an astrometric solution for the WFI image  we used 
an appropriate procedure developed at the Bologna Observatory. The resulting 
rms residuals were of the order of $\sim 0.3''$ both in RA and Dec.
By using this astrometry we were able to accurately locate
the nominal position of the MSP in the WFI image and in the HST-WFPC2 images.

Fig. 1 shows an enlargement of a $7''\times 7''$ region of the
WF4 chip, centered on the MSP position. The 
$3\sigma$ error circle, which takes into account the global error 
in the absolute positioning of the MSP, is shown. The global error
is fully dominated by the uncertainty due to the astrometric procedure.
One relatively bright star (A) has been found within the error box
of the MSP. Two additional objects (B, C) are just out-side
the error circle. In order to investigate the nature of these objects 
we performed accurate photometric analysis of the entire HST-WFPC2 
data-set retrieved from the Archive.
    
\subsection{Photometry}
The photometric reductions have been carried out using ROMAFOT (Buonanno
et al. 1983\nocite{b+83}), a package developed to perform accurate
photometry in crowded fields and specifically optimized
to handle under-sampled point spread function (PSF) as in the case of the 
HST-WF chips (Buonanno \& Iannicola 1989\nocite{bi89}).  
The standard procedure described in Ferraro et al. (1997)
was adopted in order to derive PSF-fitting instrumental magnitudes,
which were finally calibrated  using
zero-points listed by Holtzman et al. (1995). 
In particular we used a sample of median-combined images
to construct reference Color Magnitude Diagrams  (CMDs).
Figure 2 shows multiband CMDs for stars detected
in a region of $400\times400$ pixels (corresponding to
$40''\times40''$): the location of the three objects
are indicated. 

\medskip
{\plotone{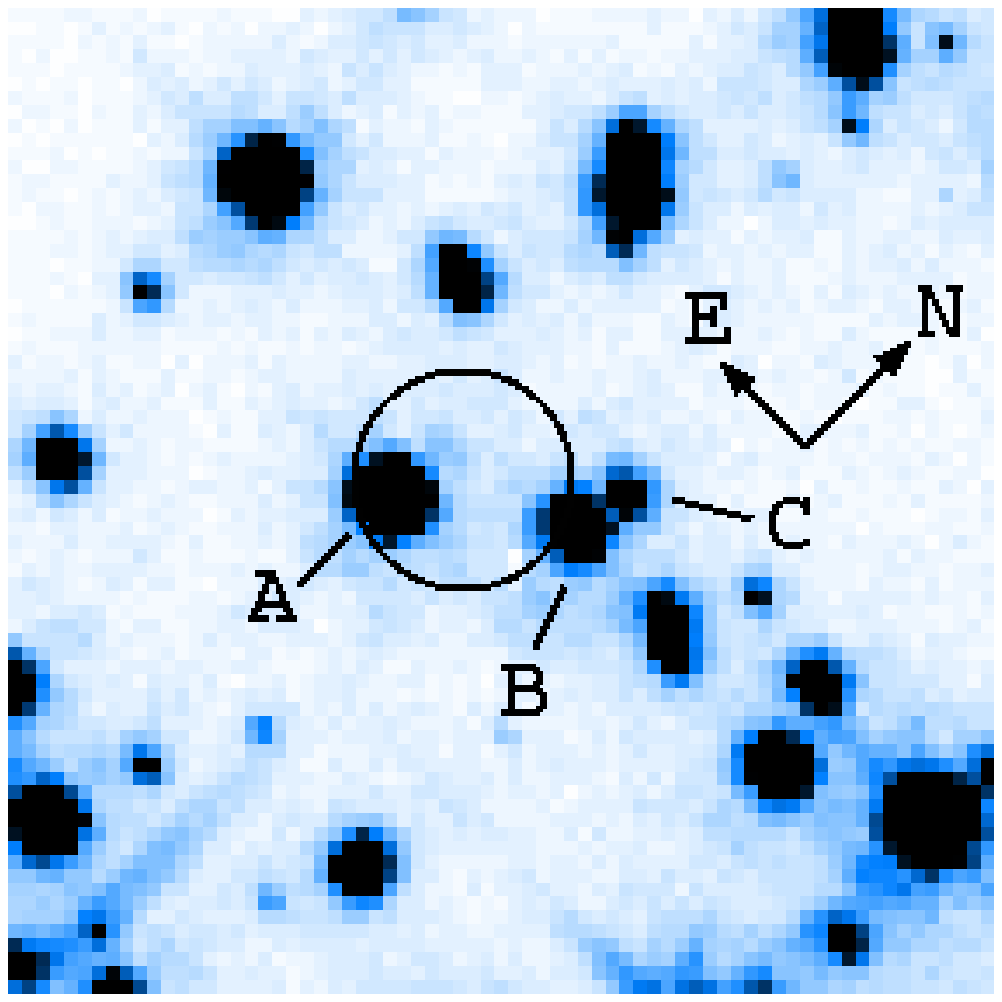}
\figcaption[f1.eps]{\footnotesize{
A portion of a median-combined F675W HST WFPC2 
image (chip WF4) of NGC 6397, centered near the MSP position.
The region covers about $7''\times 7''$.  Star A is the proposed
companion to the MSP PSR~J1740$-$5340.}}
\vskip 0.1truecm}

The result of the PSF fitting procedure for these stars have been
carefully examined by visual inspection. From this accurate 
photometric analysis we found that the bright object A
lying within the error box of the MSP has an anomalous
position in the CMD since it is located at the luminosity 
of the TO region but it has an anomalous red colour; the other two objects
(B, C) are normal Main Sequence stars. 
Individual images were instead used to check the variability
of the objects. Objects B and C show no significant time
variability compared to the measurements uncertainties. On the
other hand, object A shows a remarkable time modulation 
($\sim 0.2-0.3$ mag) on a scale of several hours. 
This object is the variable star WF4-1 proposed by Taylor \etal (2001) 
as a BY Draconis star.

\medskip
{\plotone{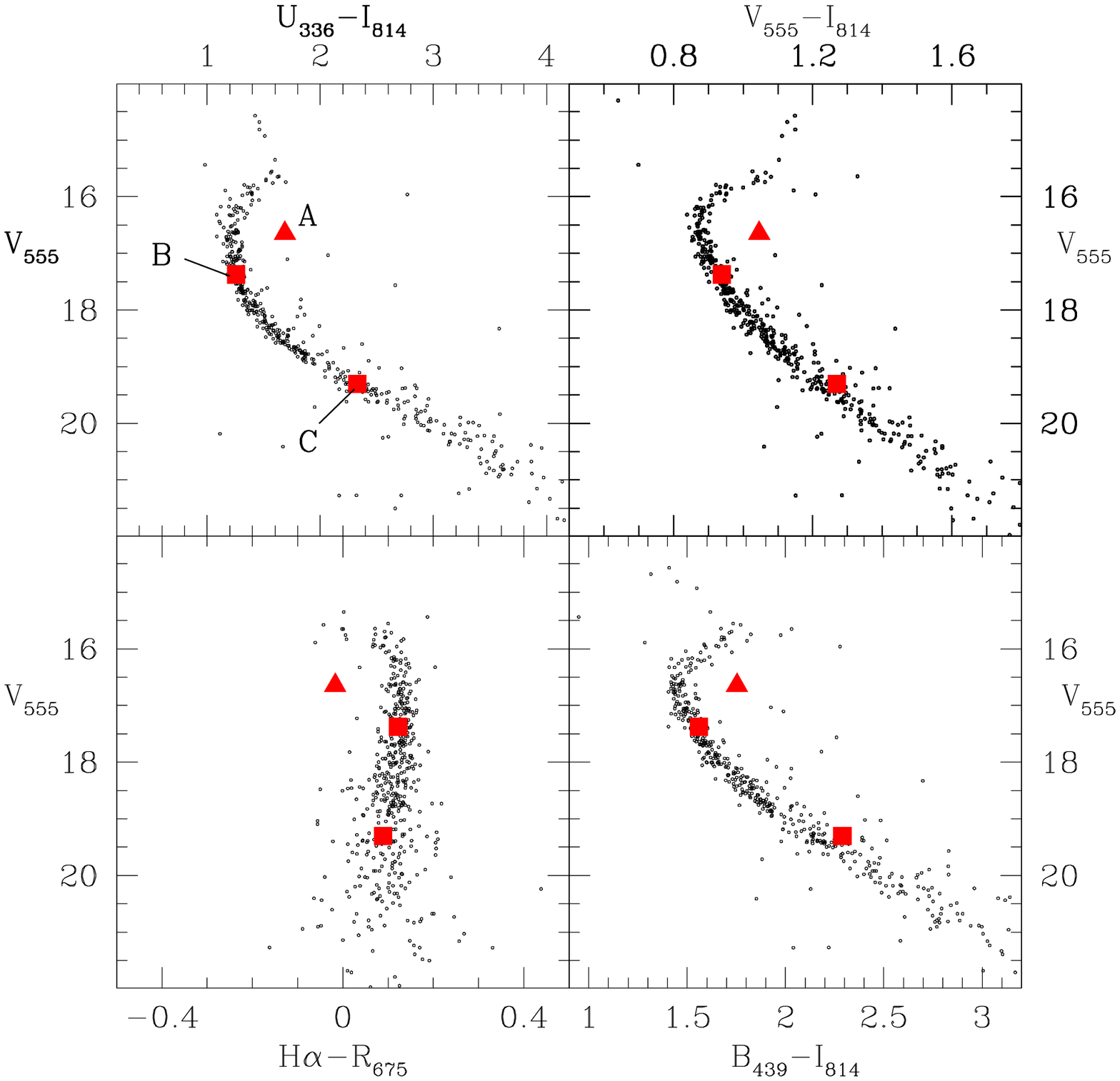}
\figcaption[f2.eps]{\footnotesize{
Multiband CMDs for stars detected in a region 
of the WF4 chip ($40''\times40''$) containing the MSP position.
The three stars found in the vicinity of the 
MSP error box are marked with different symbols 
(a triangle for star A and squares for stars B and C)  
and labeled with their names in the first panel.}}
}

\subsection{Time series}
 
In order to check the association of the time variability of
\starA to the pulsar binary motion, we have 
carried out a period search analysis.
The data available consist of four
time series in the H$\alpha$, $R_{675}, I_{814}$, and $V_{555}$ bands
taken in 1999, spanning $\sim$ 1.8 days,  and two
time series in the $B_{439}$ and $U_{336}$ bands taken in 1996, spanning
0.2 and 0.4 days, respectively.
The periodicity search was carried out
using GRATIS (GRaphycal Analyzer of TIme Series), a software
package developed at the Bologna Astronomical Observatory
(see Clementini et al. 2000, 2001).  Periods and amplitudes 
were derived for the H$\alpha$-band time series using GRATIS
$\chi^{2}$ Fourier fitting routine.  This algorithm is almost
equivalent to the Lomb-Scargle periodogram (Scargle 1982), but
has the advantage to have sensitivity also to periodicities 
whose light curve is not strictly sinusoidal, and it is more reliable
when the data span is of length comparable to the time scale 
of the stellar variation (Faulkner 1977). As already mentioned,
the data span of the H$\alpha$-band time series is $\sim$ 1.8 days, 
and the searched periodicity is 1.35 days, 
so the $\chi^{2}$ fitting method is largely preferable.

\vskip 0.3truecm
{\plotone{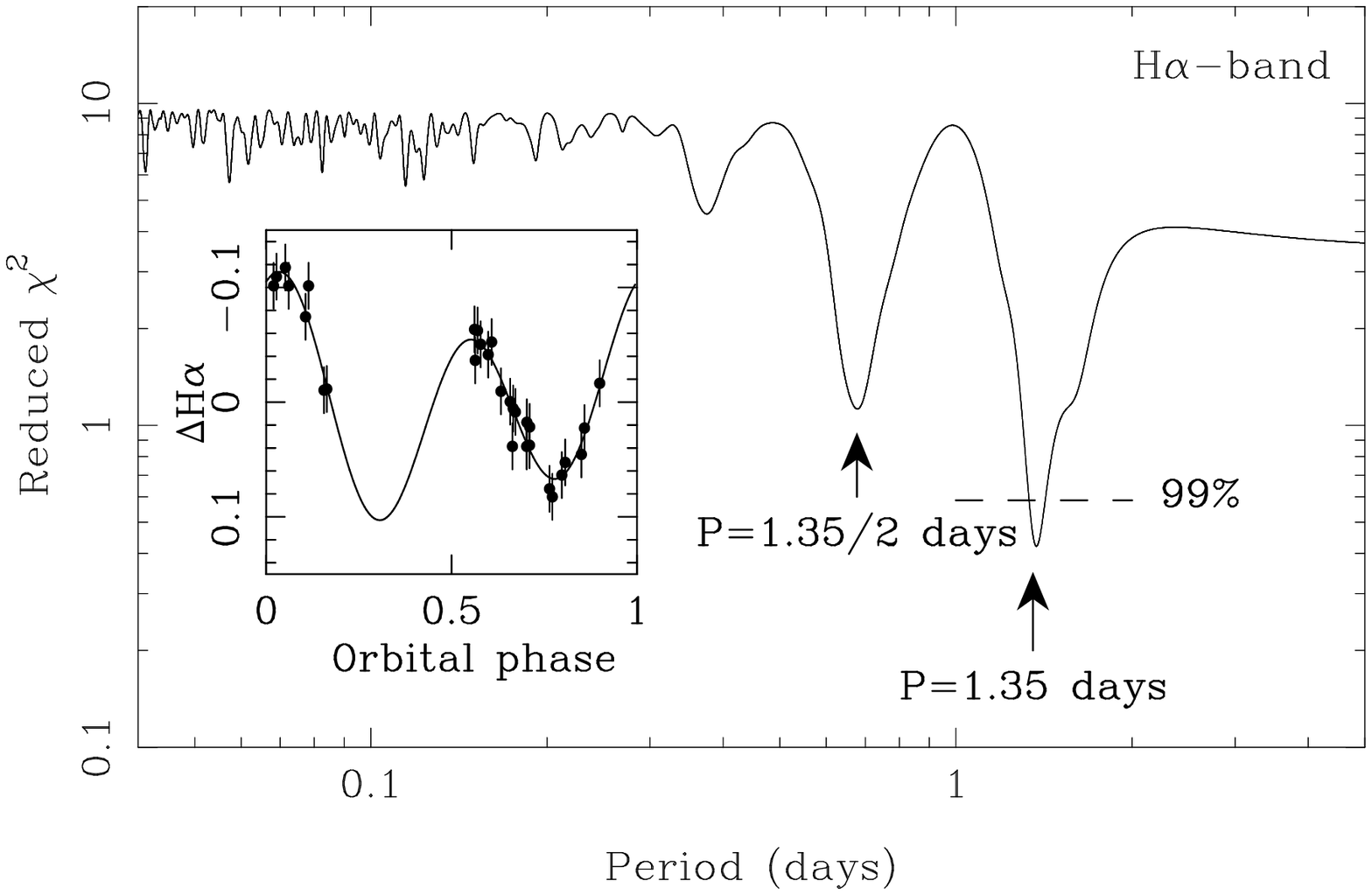}
\figcaption[f3.eps]{\footnotesize{
Reduced $\chi^{2}$ resulting from
the Fourier fitting of the H$\alpha$ data as a function 
of the modulation period. Small panel: light curve of the
H$\alpha$ data obtained adopting the period and the reference
epoch of the radio ephemeris and fitting the spectral amplitudes
of the 1$^{st}$ and 2$^{nd}$ harmonics.}
\medskip
}}

Fig. 3 shows the reduced $\chi^{2}$ resulting from
the Fourier fitting of the H$\alpha$ data as a function 
of the modulation period.  The most significant feature is 
indeed a periodicity around the predicted period of 1.35 days,
with substantial power also near the 2$^{nd}$ harmonic.  
The confidence level peak (99.6\%) corresponds to  a 
period $P$=1.37$\pm$0.05 days (consistent, within the uncertainties,
with the period quoted by Taylor \etal (2001) for the WF4-1 variable). 
The quoted uncertainty corresponds to the period range for which 
the confidence level is larger than 99\%, and it is dominated by 
the relatively short ($\sim$ 1.8 days) data span available.
          
Using another option of the GRATIS package, we have then fixed the 
period $P$ and the reference epoch $T_{0}$ to the radio ephemeris values,
and have fitted the same H$\alpha$ data  for the best spectral 
amplitudes of the 1$^{st}$ and
2$^{nd}$ harmonics.  The best-fit light curve, shown in the small
panel in Fig. 3, is not exactly what we would expect on the basis 
of a simple pulsar-irradiation model, but as we will discuss in the next
section, it could be understood in term of tidal 
distorsion effects occurring in the companion star to PSR J1740-5340. 
The time variability observed in the R, I, and V-bands at the same epochs
follows a similar pattern to that observed in the H$\alpha$-band, and the 
period search analysis produces similar results.  
Fig. 4 shows the same 1999 H$\alpha$-band data and the $U_{336}$-band 
data taken on 1996, phased using the accurate radio ephemeris.  
Remarkably, they show the minimum at the same orbital phase, giving 
further evidence that the optical modulation is indeed associated 
to the pulsar binary motion.

\vskip 0.5truecm
{\plotone{f4.eps}
\figcaption[f4.eps]{
\footnotesize{H$\alpha$-band data and U$_{336}$-band data, taken 3 yr a part,
phased using the radio ephemeris. Phase 0.0 corresponds to the 
time of the ascending node.}}}

\subsection{Is \starA the pulsar companion?}

Can we claim that \starA (WF4-1) is not a BY Draconis system but is indeed
the optical companion to the MSP? There is no
doubt that \starA is variable, and that the modulation 
period is compatible to the radio orbital period. 
However is difficult to estimate the chance occurrence probability to find 
a variable star with a period of $\sim$ 1.3$-$1.4 days in such a  
small error circle. HST observations (Taylor \etal 2001)
have discovered several variables in a WFPC2/HST field of view.  
Also, according to Taylor \etal (2001), a modulation
period of the order of $\sim$day is typical of most BY Dra systems. 

On the other hand, in two observations taken three years a part, 
we find the minima exactly at the same phase 
with respect to the precise radio orbital period, whilst   
the light curve shape of the BY Dra systems are expected 
to change, according to variations in the configuration of 
the spotted regions (Alekseev 1999)\nocite{ale99} of 
their convective envelopes. Also, the position of \starA 
in the CMD is anomalous for a BY Dra or whichever other kind of binary
system comprising two MS stars. Further support to the proposed association
derives from the detection of PSR~J1740-5340 in a 
{\sl Chandra} pointing of NGC~6397 (Grindlay \etal 2001b\nocite{ghemc2001}): 
its X-ray luminosity and color appear similar to those of the 
MSPs seen in 47 Tuc (Grindlay \etal 2001a\nocite{ghem2001})
and its positional coincidence with \starA is consistent with the
{\sl Chandra} astrometric uncertainties.

\section{Observed properties of the companion star} 

In the radio timing paper, D'Amico \etal (2001c)\nocite{dpms+01}
demonstrate that the companion star can not be the typical 
WD found in most binary MSPs. 
They propose that the companion can be a MS star acquired by exchange 
interaction in the cluster core or alternatively the same star that spun up
the MSP and that would be still overflowing its Roche lobe. Assuming that 
\starA is the pulsar companion, we here discuss these 
two hypotheses, comparing them with observed optical properties of \starA. 

In order to get some quantitative hints on the effective temperature $T_{eff}$
and the radius $R_c$ of \starA, we used the recent set of isochrones
by Silvestri \etal (1998)\nocite{svdm98} and by Vanderberg (2000)\nocite{v00}. 
By comparing the CMDs in Fig.~2 with those isochrones, 
for metallicity [Fe/H]=$-$2.00 and ages of $t=12-14$ Gyr,
(compatible with the values measured for NGC 6397), we derive
$R_{c}\sim 1.3-1.8~{\rm R_\odot}$ and $T_{eff}\sim 5500-5800$ K for \starA.

The peculiar nature of \starA can be unveiled inspecting the amplitude and the
shape of its light curves (Figure~3, 4). There are 2 other eclipsing
MSPs (PSR B1957+20 (Callanan, van Paradijs \& Rengelink 1995)\nocite{cvr95}
and PSR J2051-0827 (Stappers \etal 1999)
\nocite{svlk99}), both in the Galactic field, whose optical 
companion displays strong modulations,  
interpreted as due to the heated side of the companion entering in and
out of view according to the orbital motion.  
Similar trend, though with a much smaller degree of modulation,
is seen in 47 Tuc U$_{opt}$, the first identified MSP companion in a GC 
(Edmonds, \etal 2001). 

The light curves of \starA are completely different. We locate
the phase 0.0 at the ascending node of the MSP orbit; 
thus at the phase 0.75 we see the side of the companion
facing the pulsar. In contrast to the other known variable MSP companions, 
the light curves of \starA display there a minimum instead of a maximum
(see Figure~4). Within the limits in the orbital period coverage 
of our photometry, the best-fit light curve of Figure~3 shows two 
maxima and two minima during each binary orbit: thus, tidal distorsions 
appear the more natural responsible for this shape. 
They have been already invoked for explaining the light 
curves of the optical companions to black-hole candidates
(van der Klis, \etal 1985) and NSs (Zurita, \etal 2000). In this scenario the
maxima correspond to quadratures (phases 0.0 and 0.50), 
when the distorted star presents the longest axis of its ellipsoid 
to the observer, the minima to the conjunctions. It is easy to
recognize this trend in the insert of Figure~3. 

Whether this is the correct interpretation, it turns out in severe
constraints on the mass and the nature of Star A. The degree of
ellipsoidal variations depends roughly on \cite{r45} 
$\Delta m=k_\lambda (M_{MSP}/M_c)(R_L/a)^3F^3\sin^2i,$ 
where $M_{MSP}$ and $M_c$ are the masses of the MSP and its companion,
$R_L$ is the Roche lobe radius of the companion, $a$ is the orbital 
separation, $F$ is the ratio between the average radius of the star 
and the Roche lobe radius and $i$ is the inclination of 
the orbit. The term $k_\lambda=2.6$ accounts for limb and gravity darkening
for H$_\alpha$-radiation from our source \cite{l67}.
Given the orbital parameters of PSR J1740-5340
\cite{dpms+01}, it follows $(M_{MSP}/M_c)(R_L/a)^3\sim 0.07$ 
for all the possible companions, and thus 
$\Delta m_{H_\alpha}\la 0.2F^3\sin^2i$. 
As the observed modulation (Figure~4) is just $\sim 0.2$ mag,
we argue that only a companion almost filling its Roche lobe ($F\sim1$) 
and nearly edge-on ($i\sim 90^o$) can reproduce that. Remarkably,
these two requirements are contemporary accomplished by a
companion of mass $\la 0.25~{\rm M_\odot}$, whose Roche lobe radius
just matches the lower limit inferred for the observed radius $R_c$ of \starA. 

The light curve of a star affected only by tidally distorsion would
have the minimum at phase 0.25 less deep than that at phase 0.75.  
The reversal of this rule in the case of \starA can result from the
overheating of the side facing the pulsar.
In contrast to 47 Tuc $U_{opt}$ (having $F\sim 0.17$),
and to the companions to PSR J2051-0827 and to PSR B1957+20
(for which $F=0.5$ and $F=0.9$), \starA fills up its Roche lobe
allowing ellipsoidal variations to dominate over the thermal
modulation. 
A direct detection of both the minima would allow a 
measurement of the fraction of the impinging power from the pulsar which
goes in heating of the surface of \starA, a very interesting value for
understanding the composition of the pulsar energetic flux.

\section{Discussion}

In summary, PSR J1740-5340 appears as the first example of a MSP orbiting
a Roche lobe filling companion, whose brightness would allow unprecedented
detailed investigations, for example about the origin of this system.

A first hypothesis is that \starA is a MS star perturbed 
by the energetic flux emitted from the MSP.
The so-called {\it illumination} mechanism \cite{d95} predicts that
if the heating luminosity $L_h\la(1/4)(R_{*}/a)^2L_{irr}$ (where
$R_*$ is the star radius and $L_{irr}$ the MSP luminosity) 
is large enough, the star inflates and increases 
the effective temperature, thus modifying its photometric 
characteristics \cite{p91}.
The rotational energy loss from the MSP is 
$L_{irr}\sim 1.4\times 10^{35}~{\rm erg/s}$ \cite{dpms+01}. At the distance 
$a\sim 6.5~{\rm R_\odot}$, this corresponds to a characteristic 
temperature for the heating bath in which
the star is immersed $T_h=[1/(16\sigma\pi)L_{irr}/a^2]^{1/4}\la 4000$ K
where $\sigma$ is the constant of Stefan-Boltzmann.
We expect that the MSP flux significantly affects the companion only if
$T_h\ga T_*$ (where $T_*$ is the effective temperature of an unperturbed
MS star) and $T_*\la 4000$ K implies $\la 0.4~{\rm M_\odot}.$ 
As $T_h$ does not depend on $R_{*}$, it seems energetically difficult to
explain an increasing of $\sim 40\%$ of the effective temperature;
however, only detailed simulations (Burderi, D'Antona \& Burgay 2001, in 
preparation) of the system will allow to assess if such a low 
mass MS star of radius $\sim 0.2-0.4~{\rm R_\odot}$ can indeed be bloated 
up to $\la 1.3~{\rm R_\odot}$ and heated from $\sim 4000$ K to $\sim 5500$ K 
by the energetic flux of the MSP.

Another fascinating possibility is that PSR~J1740$-$5340 is a new-born
MSP, the first one observed just after the end of the process of
recycling. In this case \starA could have been originally a MS star 
of $1-2~{\rm M_\odot}$, whose evolution triggered mass
transfer towards the compact companion, spinning it up to millisecond periods
(Alpar, \etal 1982). Irregularities in the mass transfer rate $\dot M_{c}$ 
are common in the evolution of these systems 
(e.g. Tauris \& Savonije 1999\nocite{ts99}): even a short decreasing 
of $\dot M_{c}$ can have easily allowed 
PSR~J1740$-$5340 (having a magnetic field $\sim 8\times 10^8$ G 
and a rotational period $\la 3.65$ ms) to became source of 
relativistic particles and magnetodipole emission, whose pressure 
{\it (i)} first swept the environment of the NS,
allowing coherent radio emission to be switched on 
(Shvartsman 1970\nocite{s70}) and {\it (ii)} then kept on expelling 
the matter overflowing from the Roche lobe of \starA
(Ruderman, Shaham \& Tavani 1989\nocite{rst89}). For a wide enough 
binary system (as is the case of PSR~J1740$-$5340),
once the radio pulsar has been switched on, any subsequent restoration
of the original $\dot M_{c}$ cannot quench the radioemission 
(Burderi \etal 2001\nocite{bdm+01}). In this case
we have now a donor star still losing matter from its Roche lobe at
$\dot M_{c}\ga 5\times 10^{-11}~{\rm M_\odot/yr}$,
\cite{dpms+01} (a high mass loss rate, difficult
to explain in the model of a bloated star). At the same time,
accretion on the NS is inhibited due to the pressure exerted by the 
pulsar on the infalling matter. This strong interaction between the MSP 
flux and the plasma wind would explain also the irregularities seen 
in the radio signals from PSR~J1740$-$5340, sometimes 
showing the presence of ionized matter along the line of sight even when the
pulsar is between \starA and the observer.
The characteristic age of PSR~J1740$-$5340 ($\sim 3.5\times 10^8$ yr) 
seems indicating it is a young MSP, further supporting this scenario. 

If \starA will continue releasing matter at the present rate 
$\dot M_{c}$, PSR~J1740$-$5340 is not a candidate for 
becoming an isolated pulsar. When \starA will have shrunk
well inside its Roche lobe, the system will probably end up as
MSP+WD (or a light non degenerate companion). If \starA will undergo
a significant increasing of $\dot M_{c}$, the condition for the accretion
could be re-established and PSR~J1740$-$5340 would probably appear again as a
Low Mass X-ray Binary or as a Soft X-ray Transient (Campana, \etal 1998).

\acknowledgements

{\small We  thank P. Montegriffo for assistance with the
astrometry procedure, G. Clementini, L. Burderi and F.D'Antona 
for useful discussions and Elena Pancino for providing the WFI image.
Financial support to this research is provided by the  
Agenzia Spaziale Italiana (ASI) and the 
{\it Ministero della Universit\`a e della Ricerca Scientifica e
Tecnologica} (MURST).  The GSCII catalog was produced by the Space 
Telescope Science Institute and the Osservatorio Astronomico di Torino.
The ESO/ST-ECF Science Archive facility is a joint collaboration of 
the European Souther Observatory and the Space Telescope -
European Coordinating Facility.}

\ifx\undefined\allcaps\def\allcaps#1{#1}\fi

\end{document}